# Statistical inverse problems in active network tomography

Earl Lawrence[1,*], George Michailidis[2,*] and Vijayan N. Nair[2,*]

*Los Alamos National Laboratory and University of Michigan*

**Abstract:** The analysis of computer and communication networks gives rise to some interesting inverse problems. This paper is concerned with active network tomography where the goal is to recover information about quality-of-service (QoS) parameters at the link level from aggregate data measured on end-to-end network paths. The estimation and monitoring of QoS parameters, such as loss rates and delays, are of considerable interest to network engineers and Internet service providers. The paper provides a review of the inverse problems and recent research on inference for loss rates and delay distributions. Some new results on parametric inference for delay distributions are also developed. In addition, a real application on Internet telephony is discussed.

## 1. The inverse problems

Consider a topology with a tree structure defined as follows: $\mathcal{T} = \{\mathcal{V}, \mathcal{E}\}$ has a set of nodes $\mathcal{V}$ and a set of links or edges $\mathcal{E}$. Figure 1 shows two examples, a simple two-layer symmetric binary tree on the left and a more general four-layer tree on the right. Each member of $\mathcal{E}$ is a directed link numbered after the node at its terminus. $\mathcal{V}$ includes a (single) root node 0, a set of receiver or destination nodes $\mathcal{R}$, and a set of internal nodes $\mathcal{I}$. The internal nodes have a single incoming link and at least two outgoing links (children). The receiver nodes have a single incoming link but no children. For the tree on the right panel of Figure 1, $\mathcal{R} = \{2, 3, 6, 8, 9, 10, 11, 12, 13, 14, 15\}$ and $\mathcal{I} = \{1, 4, 5, 7\}$.

All transmissions are sent from the root (or source) node to one or more of the receiver nodes. This generates independent observations $X_k$ at all links along the paths to those receiver nodes. Let **X** denote this set of measurements. These data are not directly observable; rather we can collect only end-to-end data at the receiver nodes: $Y_r = f(\mathbf{X})$ for $r \in \mathcal{R}$. The statistical inverse problem is to *reconstruct the distributions of the link-level $X_k$s from these path-level measurements.*

Examples of $f(\cdot)$ are: $f(\mathbf{X}) = \sum_{k \in \mathcal{P}(0,r)} X_k$, $f(\mathbf{X}) = \prod_{k \in \mathcal{P}(0,r)} X_k$, and $f(\mathbf{X}) = \min_{k \in \mathcal{P}(0,r)} X_k$, and $f(\mathbf{X}) = \max_{k \in \mathcal{P}(0,r)} X_k$, where $\mathcal{P}(0, r)$ is the path between the root node 0 and the receiver node $r$. In this paper, we will be concerned only with the first two cases of $f(\cdot)$ above.

To understand the statistical issues and challenges involved, let us examine some simple examples.

*The research was supported in part by NSF Grants CCR-0325571, DMS-0204247 and DMS-0505535.
[1]Statistical Sciences Group, Los Alamos National Laboratory, Los Alamos NM 87545, USA, e-mail: earl@lanl.gov
[2]Department of Statistics, University of Michigan, Ann Arbor MI 48109, USA, e-mail: gmichail@umich.edu; vnn@umich.edu

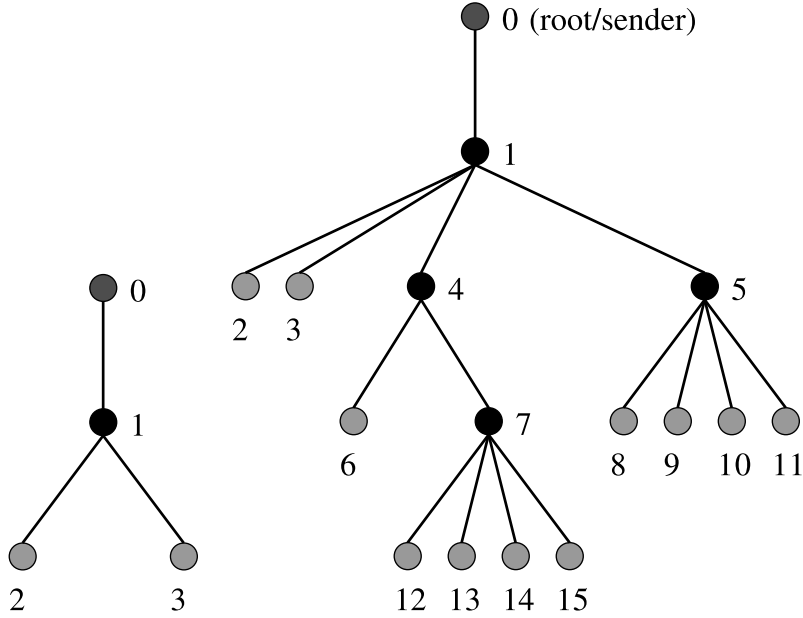

FIG 1. *Examples of tree network topologies. A binary two-layer tree is shown on the left panel and a general four-layer tree on the right panel. The path lengths from the root to nodes belonging to the same layer are the same.*

**Example 1.** Consider the two-layer binary tree on the left panel of Figure 1, and suppose the $X_k$ are binary with $P(X_k = 1) = \alpha_k$, $k = 1, 2, 3$ for the three links. Further, the root node sends transmissions to the receiver nodes *one at a time*. Take $f(a, b) = ab$. Then, the observed data are $Y_{2j} = X_{1j}X_{2j}$ for transmission $j$ and $Y_{3m} = X_{1m}X_{3m}$ for transmission $m$. They are *independent* Bernoulli with probabilities $\alpha_1\alpha_2$ and $\alpha_1\alpha_3$, respectively. Suppose we send $M$ transmissions to receiver node 2 and $N$ transmissions to receiver node 3. Let $M_1$ and $N_1$ be the respective number of "ones". Then, $M_1$ and $N_1$ are independent binomial random variables with success probabilities $\alpha_1\alpha_2$ and $\alpha_1\alpha_3$. From these data, we can estimate only $\alpha_1\alpha_2$ and $\alpha_1\alpha_3$. The individual link-level parameters $\alpha_1, \alpha_2$ and $\alpha_3$ cannot be fully recovered.

**Example 2.** Take the same two-layer binary tree with binary outcomes with $f(a, b) = ab$ as above. But now the root node sends transmissions to receiver nodes 2 and 3 *simultaneously*. In other words, the $m$-th transmission generates random variables $X_{1m}, X_{2m}$ and $X_{3m}$ on all of the links. We observe $Y_{2m} = X_{1m}X_{2m}$ and $Y_{3m} = X_{1m}X_{3m}$. The distinction from Example 1 is that the $X_{1m}$ is common to both $Y_{2m}$ and $Y_{3m}$. Now, each transmission has 4 possible outcomes: $(1, 1)$, $(1, 0)$, $(0, 1)$, $(0, 0)$ depending on whether the transmission reaches none, one, or both of the receiver nodes. If we send $N$ such transmissions to nodes 2 and 3 simultaneously, the result is a multinomial experiment with probabilities $\alpha_1\alpha_2$, $\alpha_1(1 - \alpha_2)$, $(1 - \alpha_1)\alpha_2$, and $(1 - \alpha_1)(1 - \alpha_2)$ corresponding to the four outcomes. Let $N(i, j)$ denote the number of events with outcome $(i, j)$. Then, $E[N(1, 1)] = \alpha_1\alpha_2\alpha_3$, $E[N(1, 1) + N(1, 0)] = \alpha_1\alpha_2$, and $E[N(1, 1) + N(0, 1)] = \alpha_1\alpha_3$. It is easy to see that we can estimate all the three link-level parameters from these measurements. Thus, the data transmission scheme plays an important role in this type of inverse problems.

**Example 3.** Again we have a two-layer binary tree but now $f(a, b) = a + b$. Then $Y_2 = X_1 + X_2$ and $Y_3 = X_1 + X_3$. Let $F_k$ be the distribution of the link-level random variables $X_k \in \mathcal{R}$, for $k = 1, 2, 3$. Assume, as in Example 2, that the root node sends transmissions *simultaneously* to both receivers. In this case, even with simultaneous transmission to both receivers, the link-level parameters are not always identifiable. Just take $X_k$ to be independent Normal$(\mu_k, 1)$, $k = 1, 2, 3$. Then $Y_2$ and $Y_3$ are bivariate normal with mean $\mu_1 + \mu_2$ and $\mu_1 + \mu_3$, variance 2 and correlation 1. One can see that the individual $\mu_k$ cannot be recovered from the joint distribution of $Y_2$ and $Y_3$. Additional assumptions on the distribution are needed in order to solve the inverse problem. We will revisit this issue.

**Example 4.** Consider now the more general tree on the right panel of Figure 1. Again, we send transmissions to all of the receiver nodes simultaneously. If the random variables are binary and $f(x_1, \ldots, x_p) = \prod_{j=1}^{p} x_j$, all the link-level parameters are identifiable. The same is true for a general $X_k$ with $f(x_1, \ldots, x_p) = \sum_{j=1}^{p} x_j$ under suitable conditions on the distribution of the $X_k$ (as discussed later in the paper). However, it may be "expensive" to send transmissions to all receiver nodes simultaneously. Instead, can we schedule transmissions to some judicious subsets of the receiver nodes at a time and combine the information appropriately to estimate all the link-level parameters? It is clear from Example 1 that it is not enough to send transmissions to one receiver node at a time. How should the transmission scheme be designed in order to estimate all the parameters? Are there some "good" schemes (according to some appropriate criteria)?

These examples are simple instances of issues that arise in the context of analyzing computer and communications networks and are collectively referred to as *active network tomography*. In the next section, we will describe the network application and the need for estimating quality-of-service (QoS) parameters such as loss rates and delays. Section 3 provides an overview of recent results in the literature on the design of transmission experiments and inference for loss rates and discrete delay distributions. A real application on data collected from the campus network at the University of North Carolina, Chapel Hill is used to illustrate some of the results. Section 4 develops some new results on parametric inference for delay distributions.

## 2. Active network tomography

The area of network tomography originated with the pioneering work of Vardi [14] where the term was first introduced. His work dealt with another type of inverse problem relating to origin-destination (OD) traffic matrix estimation. The OD information is important in network management, capacity planning, and provisioning. In this problem, one is interested in estimating the intensities of traffic flowing between the origin-destination pairs in the network. However, we cannot collect these data directly; rather, one places equipment at the individual nodes (routers/switches) and collects *aggregate* data on all traffic flowing through the nodes $i \in \mathcal{V}$. The goal is to recover distributions of origin-destination traffic between all pairs of nodes in the network. There has been considerable work in this area, and a summary of the developments can be found in [3].

Active network tomography, on the other hand, is concerned with the "opposite" problem of estimating link-level information from end-to-end data. One sends test probes (packets) (active probing) from a source to one or more receiver nodes on the periphery of the network and gets end-to-end path-level data on losses and delays. One then has to solve the inverse problem of reconstructing link-level loss and delay information from the end-to-end data. The specific goal is to estimate QoS parameters such as loss rates and delays at the link level. The reason for probing the network from the outside is that Internet service providers or other interested parties often do not have access to the internal nodes of the network (which may be owned by a third party). Nevertheless, they have to assess QoS of the links over which they are providing service. Active tomography offers a convenient approach by probing the network from nodes located on the periphery.

The probing and data collection are done with dedicated instruments at the root node and receiver nodes. These packets can be sent to one receiver at a time (unicast transmission scheme) or to a specified subset of receivers (multicast scheme). Some networks have turned off the multicast scheme for security reasons. In this case, one sends unicast packets to several receivers spaced closely in time with the goal of trying to mimic the multicast scheme.

What causes losses and delays of packets over the network? When a packet arrives at a node, it joins a queue of incoming packets. If the buffer is full, the packet is dropped, i.e., lost. Depending on the protocol, the packet may or may not be resent. Packets also encounter delays along the path, primarily due to the queueing process above.

In the case of losses, the binary outcome $X_k = 0$ or 1 indicates whether the packet is lost (dropped) or not. In terms of the examples in Section 1, $f(x_1, \ldots, x_k) = \prod_k x_k$, and the end-to-end loss $Y = \prod_{k \in \mathcal{P}(0,r)} X_k = 1$ if the packet transmitted along the path $\mathcal{P}(0, r)$ reached the receiver node $r$ and zero otherwise. For delays, $f(x_1, \ldots, x_K) = \sum_k x_k$, and the end-to-end observation is $Y = \sum_{k \in \mathcal{P}(0,r)} X_k$, the path-level delay.

The physical topology of a network is usually complicated. But the logical topology with a single source node can often be represented as a tree. For example, the left panel of Figure 2 shows the physical topology of a subnetwork at the campus of the University of North Carolina at Chapel Hill. The right panel shows the corresponding logical topology, which is a tree with a directed flow. We will revisit this network later in the paper. It is possible to deal with topologies with multiple sources, other kinds of transmission schemes (two-way flows), and so on. But for simplicity, we will restrict attention to the tree structures in this paper.

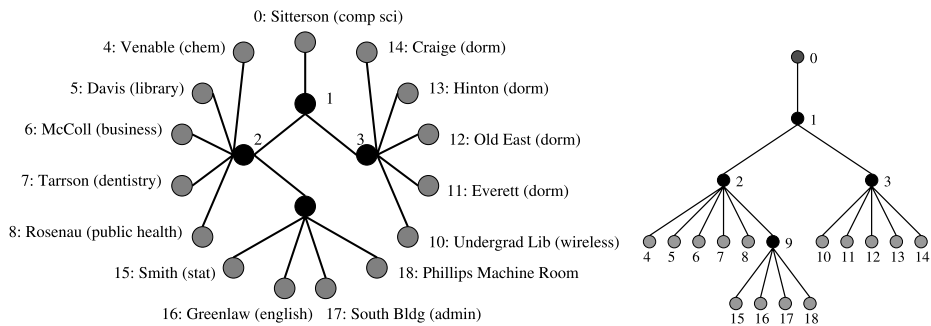

FIG 2. *Left panel: Schematic of the UNC network; Right panel: Logical topology of the UNC network.*

## 3. Literature review of loss and discrete delay inference

Most of the results in the literature on active tomography have been developed under the assumption that the loss rates and delay distributions are temporally homogeneous and are independent across links. We will also use this framework. The assumption of temporal homogeneity is reasonable as the probing experiments are done within the order of minutes. The assumption of independence across links is less likely to hold. However, the nature of the dependence will vary from network to network, and it is difficult to obtain general results.

### *3.1. Design of probing experiments*

We noted in Example 1 that the link-level parameters are not identifiable under the unicast transmission scheme (sending probes to one receiver at a time). The multicast scheme, which sends packers to *all* the receivers in the network simultaneously, addresses this problem for loss rates and, under some additional conditions, for delay distributions as well.

However, this scheme has a number of drawbacks. It creates more traffic than necessary for estimating the link-level parameters. Also, the data generated are very high-dimensional. For example, in a binary symmetric tree with $L$ layers, there are $R = 2^L - 1$ receiver nodes. A multicast scheme for measuring loss rates results in a multinomial experiment with $2^R$ possible outcomes. This is a large number even for moderately sized trees. The most important drawback, however, is that it is inflexible and does not allow investigation of subnetworks using different intensities and at different times. In practice, one may want to probe sensitive parts of the network as lightly as necessary to avoid disturbance. So there is a need for more flexible probing experiments. As pointed out in Example 4, this raises interesting issues on how to design the probing experiments.

A class of flexible probing experiments, called *flexicast* experiments, were introduced and studied in Xi et al. [17] and Lawrence et al. [8]. This consists of a combination of schemes for different values of $k$ with each scheme aimed at studying a subnetwork. However, each of the scheme by itself will not necessarily allow us to estimate the link-level parameters of that subnetwork. The data have to be combined across the various k-cast schemes to estimate the link-level parameters.

To illustrate the ideas, consider the network on the right panel in Figure 1. The multicast scheme sends probes simultaneously to $\{\langle 2, 3, 6, 8, 9, 10, 11, 12, 13, 14, 15\rangle\}$. Two possible flexicast experiments are:

(1) $$\{\langle 2, 3\rangle, \langle 6, 12\rangle, \langle 13, 14\rangle, \langle 8, 15\rangle, \langle 9, 10\rangle, \langle 11\rangle\}$$

and

(2) $$\{\langle 2, 3\rangle, \langle 6\rangle, \langle 12, 13, 14, 15\rangle, \langle 8, 9, 10, 11\rangle\}.$$

The former consists of only bicast (two receiver nodes at a time) and unicast schemes. Intuitively, the latter scheme appears to more "efficient" but we will see shortly that it does not allow one to estimate all the link-level parameters.

A full multicast scheme for this tree will result in 11-tuples or 11-dimensional data. The first flexicast experiment using pairs and singletons can cover the whole tree with five pairs and one singleton. The resulting data are considerably less complex in terms of processing and computations for inference. This advantage is particularly important for trees with many layers.

Of course, not all flexicast experiments will permit estimation of the link-level parameters. To discuss the technical issues associated with the identifiability problem, consider first the notion of a splitting node. For a $k$-cast scheme, an internal node is a splitting node if the scheme splits at that node. For example, for the tree on the right panel of Figure 1, the bicast scheme $\{\langle 6, 12 \rangle\}$ splits at node 4. Xi et al. [17] showed that the following conditions are necessary and sufficient for identifiability of link-level loss rates: (a) all receiver nodes are covered; and (b) every internal node in the tree is a splitting node for some k-cast scheme in the flexicast experiment. Lawrence et al. [8] studied the delay problem and showed that the same conditions are also necessary and sufficient for estimating delay distributions provided the distributions are discrete. The case where the delay distributions are not discrete is discussed in the next section.

Consider again the flexicast schemes in equations (1) and (2) for the tree on the right panel in Figure 1. The first one based on a collection of bicast and unicast schemes satisfies the conditions. For the second one, none of the k-cast schemes split at node 4.

There are many flexicast experiments that satisfy the identifiability requirements, and the choice among these has to be based on other criteria. Experiments based on just bicast and unicasts have minimal data complexity – just 1- and 2-dimensional outcomes. However, these provide information on just first and second-order dependencies and will be less efficient (in a statistical sense) to k-cast schemes with higher values of $k$. In particular, the full mulitcast scheme will be most efficient in this sense. So the overall choice of the flexicast experiment has to be a compromise between statistical efficiency and flexibility including the ability to adapt over time to accommodate changes in network conditions.

## 3.2. Inference for loss rates

Inference for loss rates was first studied in Cáceres et al. [2] for the multicast scheme. A recent, up-to-date list of references can be found in Xi et al. [17] who developed MLEs based on the EM algorithm for flexicast experiments. We provide next a brief review of these results.

Each $k$-cast scheme in a flexicast experiment is a $k$-dimensional multinomial experiment. Specifically, each outcome is of the form $\{Z_{r_1}, \ldots, Z_{r_k}\}$ where $Z_{r_j} = 1$ or 0 depending on whether the probe reached receiver node $r_j$ or not. Let $N_{(r_1,\ldots,r_k)}$ denote the number of outcomes corresponding to this event, and let $\gamma_{(r_1,\ldots,r_k)}$ be the probability of this event. Then the log-likelihood for the $k$-cast scheme is proportional to $\gamma_{(r_1,\ldots,r_k)} \log(N_{(r_1,\ldots,r_k)})$. The overall log-likelihood is just the sum of the log-likelihoods for these individual experiments. However, the $\gamma_{(r_1,\ldots,r_k)}$ are complicated functions of $\alpha_k$, the link-level loss rates, so one has to use numerical methods to obtain the MLEs.

The EM algorithm is a natural approach for computing the MLEs and has been used extensively in network tomography applications (see [3, 5, 16]). The structure of the EM-algorithm for general flexicast experiments was developed in Xi et al. [17]. While the E-step can be complex for arbitrary collections of $k$-cast schemes, it simplifies for flexicast experiments comprised of bicast and unicast schemes as seen below.

Let $s_b$ be the splitting node for bicast pair $b = \langle i_b, j_b \rangle$. Then, $\pi(0, s_b)$, $\pi(s_b, i_b)$ and $\pi(s_b, j_b)$, the three path probabilities for this bicast pair are products of the $\alpha_k$. Starting with an initial value $\vec{\alpha}^{(0)}$ let $\vec{\alpha}^{(k)}$ be the value after the $k$-th iteration. Then, we can write the $(k+1)$-th iteration of the E-step as follows:

**E-step:**

1. For each bicast pair:

   (a) Use $\vec{\alpha}^{(k)}$ to obtain the updated path probabilities $\pi^{(k)}(0, s_b)$, $\pi^{(k)}(s_b, i_b)$, $\pi^{(k)}(s_b, j_b)$ and $\gamma_{00}^{b\,(k)}$.

   (b) For each node $\ell \in \mathcal{P}(0, s_b) \cup \mathcal{P}(s_b, i_b) \cup \mathcal{P}(s_b, j_b)$, compute $V_{\ell,b}^{(k+1)} = E_{\vec{\alpha}^{(k)}}[V_\ell | \mathcal{N}_b]$, where $\mathcal{N}_b = \{N_{00}^b, N_{01}^b, N_{10}^b, N_{11}^b\}$ are the collected counts of the four possible outcomes, as follows.
   For node $\ell \in \mathcal{P}(0, s_b)$,

   $$V_{\ell,b}^{(k+1)} = N^b - N_{00}^b \frac{1 - \alpha_\ell^{(k)}}{\gamma_{00}^{b\,(k)}}.$$

   For link $\ell \in \mathcal{P}(s_b, i_b)$,

   $$V_{\ell,b}^{(k+1)} = N^b - N_{01}^b \times \frac{1 - \alpha_\ell^{(k)}}{1 - \pi^{(k)}(s_b, i_b)} - N_{00}^b \frac{(1 - \alpha_\ell^{(k)})(1 - \pi^{(k)}(0, j_b))}{\gamma_{00}^{b\,(k)}}.$$

   For link $\ell \in \mathcal{P}(s_b, j_b)$,

   $$V_{\ell,b}^{(k+1)} = N^b - N_{10}^b \times \frac{1 - \alpha_\ell^{(k)}}{1 - \pi^{(k)}(s_b, j_b)} - N_{00}^b \frac{(1 - \alpha_\ell^{(k)})(1 - \pi^{(k)}(0, i_b))}{\gamma_{00}^{b\,(k)}}.$$

2. Unicast schemes: Let node $\ell \in \mathcal{P}(0, u)$ for a unicast transmission to receiver node $u$, and compute

$$V_{\ell,u}^{(k+1)} = N^u - N_0^u \times \frac{1 - \alpha_\ell^{(k)}}{1 - \pi^{(k)}(0, u)}$$

**M-step:** The $(k+1)$-th update for the M-step is simply

$$\alpha_\ell^{(k+1)} = \frac{\sum_{b \in \mathcal{B}_\ell} V_{\ell,b}^{(k+1)} + \sum_{u \in \mathcal{U}_\ell} V_{\ell,u}^{(k+1)}}{\sum_{b \in \mathcal{B}_\ell} N^b + \sum_{u \in \mathcal{U}_\ell} N^u}$$

where $\mathcal{B}_\ell$ is the set of bicast pairs that includes the node $\ell$ in its path and $\mathcal{U}_\ell$ is the set of all unicast schemes that includes node $\ell$ in its path.

In our experience, the EM algorithm works reasonably well for small to moderate networks when used with a flexicast experiment that consists of a collection of bicast and unicast schemes. For large networks, however, it becomes computationally intractable. In on-going work, we are developing a class of fast estimation methods based on least-squares methods and are studying their application to on-line monitoring of network performance.

### 3.3. Inference for discrete delay distributions

For the delay problem, let $X_k$ denote the (unobservable) delay on link $k$, and let the cumulative delay accumulated from the root node to the receiver node $r$ be $Y_r = \sum_{k \in \mathcal{P}(0,r)} X_k$. Here $\mathcal{P}(0, r)$ denotes the path from node $0$ to node $r$. The observed data are end-to-end delays consisting of $Y_r$ for all the receiver nodes.

Most of the papers on delay inference assume a discrete delay distribution. Specifically, if $q$ denotes the universal bin size, $X_k \in \{0, q, 2q, \ldots, bq\}$ is the discretized delay on link $k$ and $bq$ is the maximum delay. Let $\alpha_k(i) = \mathrm{P}\{X_k = iq\}$. The inference problem then reduces to estimating the parameters $\alpha_k(i)$ for $k \in \mathcal{E}$ and $i$ in $\{0, 1, \ldots, b\}$ using the end-to-end data $Y_r$. Lo Presti et al. [10] developed a fast, heuristic algorithm for estimating the link delays. Liang and Yu [9] developed a pseudo-likelihood estimation method. Nonparametric maximum likelihood estimation under the above setting was investigated in Tsang et al. [13] and Lawrence et al. [8]. Shih and Hero [12] examined inference under mixture models. See also Zhang [18] for a more general discussion of the deconvolution problem.

We discuss nonparametric MLE with discrete delays in more detail. Let $\vec{\alpha}_k = [\alpha_k(0), \alpha_k(1), \ldots \alpha_k(b)]'$ and let $\vec{\alpha} = [\vec{\alpha}_0', \vec{\alpha}_1', \ldots, \vec{\alpha}_{|\mathcal{E}|}']'$. The observed end-to-end measurements consist of the number of times each possible outcome $\vec{y}$ was observed from the set of outcomes $\mathcal{Y}^c$ for a given scheme $c$. Let $N_{\vec{y}}^c$ denote these counts. These are distributed as multinomial random variables with corresponding path-level probabilities $\gamma_c(\vec{y}; \vec{\alpha})$. So the log-likelihood is given by

$$l(\vec{\alpha}; \mathbf{Y}) = \sum_{c \in \mathcal{C}} \sum_{\vec{y} \in \mathcal{Y}^c} N_{\vec{y}}^c \log[\gamma_c(\vec{y}; \vec{\alpha})].$$

This cannot be maximized easily, and one has to resort to numerical methods.

Again, the EM algorithm is a reasonable technique for computing the MLEs. See [7] for multicast schemes and [8] for inference with flexicast experiments. However, the complexity of the EM algorithm, in particular computing conditional expectations of the internal link delays for each bin, is prohibitive for all but fairly small-sized networks. To deal with larger networks, [8] developed a *grafting* method which fits "local" EMs to the subtrees defined by the $k$-cast schemes and then combines the estimates through a fixed point algorithm. This hybrid algorithm is fast and has reasonable statistical efficiency compared to the full MLE.

For bicast schemes, the resulting algorithm has third-order polynomial complexity, a substantial improvement over the full bicast MLE. The heuristic algorithm in [10] is based on solving higher order polynomials and is much faster. However, it uses only part of the data and is quite inefficient. The pseudo-likelihood method of [9] uses only data from all pairs of probes in the multicast experiment. This is similar in spirit to a flexicast experiment comprised of only bicast schemes, although in this setting the schemes would be independent. The computational performance of the pseudo-likelihood method is faster than the MLE based on the full multicast. It is comparable to doing a full EM based on data from all possible bicast schemes. This will still not scale up well to very large trees as it includes all possible bicasts which can involve a large number of schemes. Furthermore, using the full MLE combining the results across all schemes is computationally intensive. The flexicast experiments, on the other hand, are typically based on a much smaller number of schemes (eve if one restricts attention to bicasts). Further, the grafting algorithm is much faster for combining the results across the schemes.

### *3.4. Application to the UNC network*

We use a real example to demonstrate how the results from active tomography are used. The example deals with estimating the QoS of the campus network at the University of North Carolina at Chapel Hill and assessing its capabilities for Voice-Over-IP readiness.

This network has 15 endpoints which were organized into the tree shown in Figure 2. Node 1 is the main campus router and it connects to the university gateway. Nodes 2, 3, and 9 are also large routers responsible for different portions of the campus. The accessible nodes are all located in dorms and other university buildings. The root node of the tree was Sitterson Hall which houses the computer science department. The network was probed in pairs using the following flexicast experiment: $\{\langle 4,5\rangle, \langle 6,7\rangle, \langle 8,10\rangle, \langle 11,12\rangle, \langle 13,14\rangle, \langle 15,16\rangle, \langle 17,18\rangle\}$. A single probing session consisted of two passes through the collection of experiments sending about 500 probes to each pair in a single pass. The experiment was conducted over the course of several days in order to evaluate both the network and the methodology. We have collected extensive data but show only selected results for illustrative purposes.

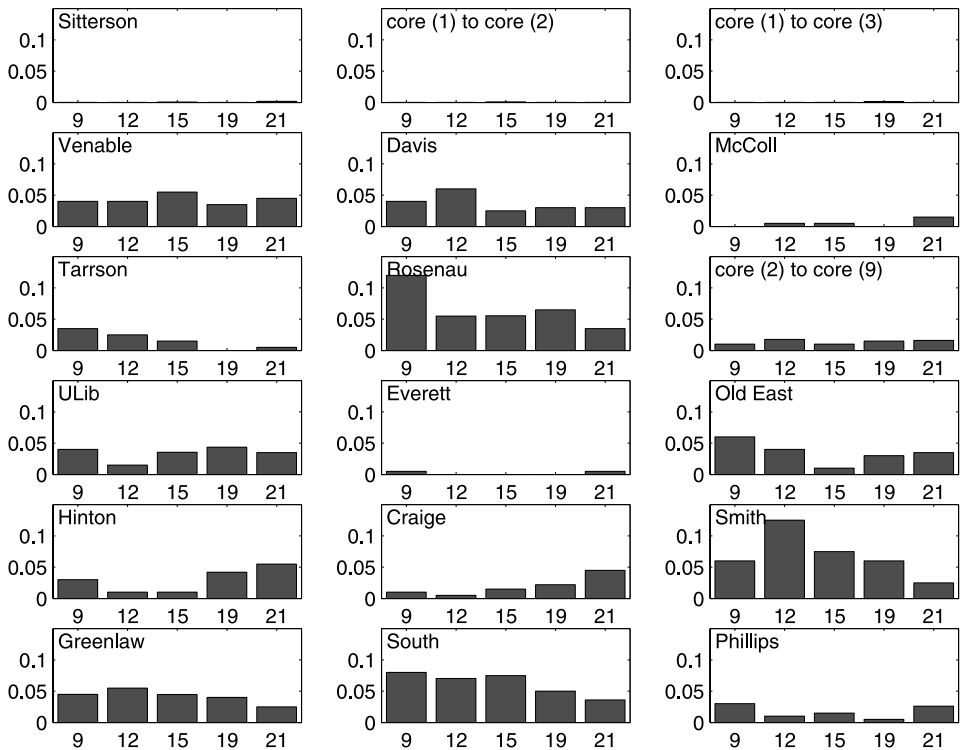

Fig 3. *Probability of large delay on 3/7/2005.*

The data presented here were collected at 9:00 a.m., 12:00 p.m., 3:00 p.m., 6:00 p.m., and 9:00 p.m. on March 1 and 17 of 2005. March 17 was during spring break. For both days, we chose a bin size of $q = .0001s$ to assess occurrences of large delays on the network. The large bin size also allowed us to use the full MLE to estimate the delay distributions. Figures 3 and 4 provide a picture of the probability of large delay (larger than a specified threshold) throughout the course of the day.

From Figure 3, we see that many buildings (Venable, Davis, Rosenau, Smith, Greenlaw, and South) show a typical diurnal pattern. These buildings are either administrative or departmental building; so the majority of users follow a regular 9 to 5 schedule. Other buildings are either more uniform throughout the day or even more activity at night. Hinton, for example, is a large freshman dorm and thus the drop during the day and increase at night are expected as the residents return from classes and other activities in the evening.

A comparison of Figures 3 and 4 shows the difference in dorm activity before and during spring break. Everett, Old East, Hinton, and Craige are dorms. The data collected during spring break reveals almost no large delays in three out of four of these buildings. This is of course to be expected. The Hinton dorm is especially interesting, since it experienced very little congestion over the break, but a significant increase to pre-break levels on the first day after the break (post-break results are not shown here).

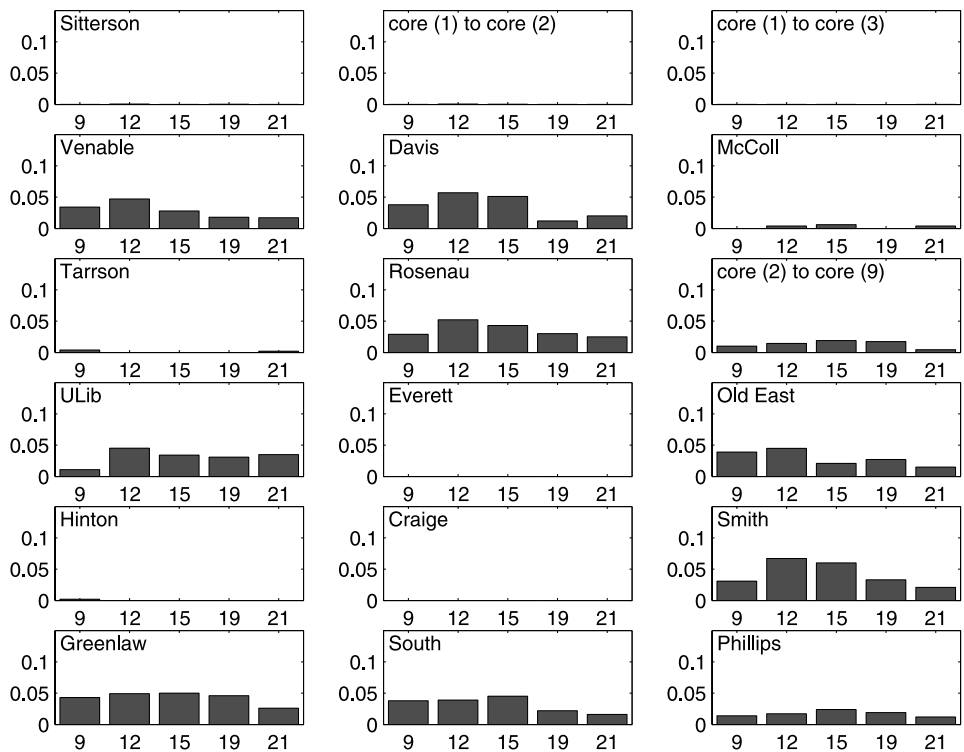

FIG 4. *Probability of large delay on 3/17/2005.*

As a consequence of this study, it became clear that many of the building links require upgrades in order to support delay-sensitive applications such as VoIP. Some of the departmental and administration buildings (Smith and South) already have large delays even without additional VoIP traffic.

## 4. Parametric inference for delay distributions

This section develops some new results on parametric inference for delay distributions. We start with a framework that includes two components: a zero delay and a (non-zero) finite delay. Specifically, let $X_k$ be the delay on link $k$, and suppose

$$(3) \qquad X_k \sim p_k \delta_{\{0\}} + (1 - p_k) F(x; \theta_k).$$

Here we assume that $F(x; \theta_k)$ does not give any mass to zero, for all $k$. So, a successful transmission (finite delay or no loss) experiences an empty queue (no delay) with probability $p_k$ and has some non-zero delay that is distributed according to a parametric distribution $F(\cdot)$ indexed by $\theta_k$ with probability $1 - p_k$.

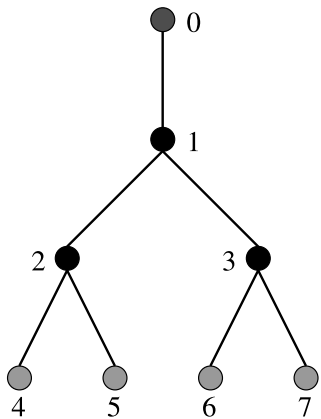

FIG 5. *Three-layer, binary tree*

### 4.1. Identifiability

The basic issue for delay distributions is the one posed in Example 2 in the introductory section, viz., whether the parameters of a simple two-layer tree (left panel of Figure 1) are estimable from probes sent simultaneously to both receivers. If this holds, then the result extends readily to general flexicast experiments that satisfy the conditions in Section 3.1 (using the arguments in [8]). We discuss the details briefly. See also [4, 6] for a general discussion of identifiability issues.

We consider two cases:

Case 1: If $p_k > 0$ for all $k$, no additional assumptions on the distribution $F(\cdot)$ are needed. All the link-level delay parameters ($p_k$ and $\theta_k$) are identifiable using flexicast experiments provided they satisfy the conditions in Section 3.1: a) every receiver node is covered and b) every internal node is a splitting node for some sub-experiment.

To see this, consider the two-layer tree on the left panel of Figure 1. Condition on the subset of data with $Y_{2,m} = 0$ and $Y_{3,m} > 0$ for probes $m = 1, \ldots, M$. Now, $Y_{2,m} = 0$ implies that both of the internal links $X_{1,m}$ and $X_{2,m}$ had zero delay, so $Y_{3,m} = X_{3,m}$. So we can use this subset of $Y_{3,m}$ to estimate $F(x; \theta_3)$. A similar argument can be used to estimate $F(x; \theta_2)$ using the subset of $Y_{3,m} > 0$ and $Y_{2,m} = 0$. Once these two distributions are estimated, we can easily estimate $F(x; \theta_1)$.

Case 2: If $p_k = 0$, then we need additional assumptions on the delay distributions $F(x; \theta_k)$. As we noted in Example 2, the means of the normal distributions are not identifiable. If the moments of order two and higher depend on the first moment, they will provide additional information for estimating the parameters. One such example is when the variance is a function of the mean (as is the case with exponential, gamma, log-normal, and Weibull distributions).

**Example 5.** We consider here a more general situation with the three-layer binary, symmetric tree shown in Figure 5. Let the delay on link $k$ be distributed Gamma($\alpha_k, \beta_k$). Suppose we use the flexicast probing experiment $\{\langle 4,5 \rangle, \langle 5,6 \rangle, \langle 6,$

7⟩}. The covariances yield the following moment equations:

$$\text{Cov}(Y_5^{\langle 5,6\rangle}, Y_6^{\langle 5,6\rangle}) = \alpha_1 \beta_1^2,$$
$$\text{Cov}(Y_4^{\langle 4,5\rangle}, Y_5^{\langle 4,5\rangle}) - \text{Cov}(Y_5^{\langle 5,6\rangle}, Y_6^{\langle 5,6\rangle}) = \alpha_2 \beta_2^2,$$
$$\text{Cov}(Y_6^{\langle 6,7\rangle}, Y_7^{\langle 6,7\rangle}) - \text{Cov}(Y_5^{\langle 5,6\rangle}, Y_6^{\langle 5,6\rangle}) = \alpha_3 \beta_3^2.$$

Let $\text{E}(Y_r) = \nu_r$. We also get the following equations based upon third moments:

$$\text{E}(Y_5^{\langle 5,6\rangle} - \nu_5)^2 (Y_6^{\langle 5,6\rangle} - \nu_6) = 2\alpha_1 \beta_1^3,$$
$$\text{E}(Y_4^{\langle 4,5\rangle} - \nu_4)^2 (Y_5^{\langle 4,5\rangle} - \nu_5) - \text{E}(Y_5^{\langle 5,6\rangle} - \nu_5)^2 (Y_6^{\langle 5,6\rangle} - \nu_6) = 2\alpha_2 \beta_2^3,$$
$$\text{E}(Y_6^{\langle 6,7\rangle} - \nu_6)^2 (Y_7^{\langle 6,7\rangle} - \nu_7) - \text{E}(Y_5^{\langle 5,6\rangle} - \nu_5)^2 (Y_6^{\langle 5,6\rangle} - \nu_6) = 2\alpha_3 \beta_3^3.$$

The corresponding sample moments can be used to estimate the terms on the left. Then, estimators for $\alpha_1, \beta_1, \alpha_2, \beta_2, \alpha_3$, and $\beta_3$ can be obtained by rearranging the above equations.

The parameters for the receiver links can be estimated with just the first moments. For example, the equations for link 4 are:

$$\text{E}(Y_4) = \alpha_1 \beta_1 + \alpha_2 \beta_2 + \alpha_4 \beta_4,$$
$$\text{Var}(Y_4) = \alpha_1 \beta_1^2 + \alpha_2 \beta_2^2 + \alpha_4 \beta_4^2.$$

The unknown parameters are easily obtained from the observed values on the left and the estimated parameters on the right.

### 4.2. Maximum likelihood estimation

It turns out that $p_k$, the probability of zero delay, can be estimated using methods analogous to those for loss rate discussed in Section 3.2. Recall that a zero delay will be observed at the receiver node if and only if there is zero delay at every link. On the other hand, a non-zero delay at the receiver link may include zero delays at some links, so we have to "recover" this information from the aggregate level data. But this is equivalent to the problem with of losses. A packet received at the receiver node implies "success" at all the links. A packet not received at the receiver node involves a combination of successes and losses, with at least one loss. Thus, we can use the data with zero-delays and positive delays in an analogous manner to estimate the zero-delay probabilities $p_k$.

To simplify matters, therefore, we will focus on parametric estimation of $F_k(x; \theta_k)$ assuming that $p_k = 0$. Let us consider some simple examples with the two-layer tree with two receivers in the left panel of Figure 1 and with exponential and gamma distributions for delays. The gamma family is closed under convolution if the scale parameters are the same, so the distribution of the end-to-end delays belong to the same family as the link-level delays. Even for these simple cases, we will see that the MLE computations are intractable.

**a) Exponential Distributions:** Suppose the delay distribution on each link is exponential with parameter $\lambda_k$. Further, we send $N$ probes to both receivers $\langle 2, 3 \rangle$ simultaneously. The log-likelihood function is

$$l(\vec{\lambda}; \mathbf{Y}) = N\log(\lambda_1) + N\log(\lambda_2) + N\log(\lambda_3) - N\log(\lambda_1 - \lambda_2 - \lambda_3) \qquad (4)$$
$$- \lambda_2 \sum_{i=1}^{N} y_{i,2} - \lambda_3 \sum_{i=1}^{N} y_{i,3}$$
$$+ \sum_{i=1}^{N} \log[1 - \exp\{-(\lambda_1 - \lambda_2 - \lambda_3)\min(y_{i,2}, y_{i,3})\}]$$

There is no analytic solution to maximize this equation over $\vec{\lambda}$, so one would have to use an iterative technique, such as EM or Newton-Raphson, to find the MLEs even in this simple case.

We examine the details for the EM-algorithm. The exponential distribution is a member of the exponential family, so the (unobserved) sufficient statistics are the total link-level delays $\sum_{i=1}^{n} X_{i,1}$, $\sum_{i=1}^{n} X_{i,2}$, and $\sum_{i=1}^{n} X_{i,3}$. Since $X_{i,2} = Y_{i,2} - X_{i,1}$ and $X_{i,3} = Y_{i,3} - X_{i,1}$, we need to compute only the conditional expected values of $\sum_{i=1}^{n} X_{i,1}$ in the E-step. The conditional distribution $[X_1|Y_2 = a, Y_3 = b]$ has density

$$g(x) = \frac{exp\{-(\lambda_1 - \lambda_2 - \lambda_3)x\}}{C(a,b; \lambda_1, \lambda_2, \lambda_3)}, \quad 0 < x < a \wedge b, \qquad (5)$$

where $a \wedge b = \min(a,b)$ and the constant of proportionality is

$$C(a,b; \lambda_1, \lambda_2, \lambda_3) = \frac{1 - \exp\{-(\lambda_1 - \lambda_2 - \lambda_3)(a \wedge b)\}}{\lambda_1 - \lambda_2 - \lambda_3}.$$

Now $\int_0^{a \wedge b} xg(x)dx$ is an incomplete gamma function and one can compute the expected value $\sum_{i=1}^{n} E(X_{i,1}|Y_{i,2}, Y_{i,3})$ as a ratio of the incomplete gamma function and the constant $C(a,b; \lambda_1, \lambda_2, \lambda_3)$. Thus, the MLEs of the $\lambda_k$ can be computed without too much trouble in this simple two-layer binary case.

How well does this extend to more general cases? Suppose we have a three layer binary tree (Figure 5), and we use bicast schemes $\langle 4,5 \rangle$, $\langle 6,7 \rangle$, and $\langle 5,6 \rangle$. Consider the scheme $\langle 4,5 \rangle$ which splits at node 2. We can try and mimic the computations for the two-layer tree above. However, we have to consider the combined path $\mathcal{P}_{(0,2)}$ whose delay is the sum of delays for links 1 and 2. The exponential distribution is not closed under convolution, so the distribution is now more complex. The details for more general trees will depend on the number of links involved before-and-after the splitting node. The problem is even more complex for multicast schemes with multiple splitting nodes. We see that the MLE computations are complicated even for simple exponential distributions.

**b) Gamma Distributions**: Gamma distributions with same scale parameter are closed under convolution, i.e., the path delays which are sums of link-level delays are still gamma. Specifically, let $X_k \sim \text{Gamma}(\alpha_k, \beta)$ and independent across $k$ for $k \in \mathcal{E}$. We start with the simple two-layer binary tree. Then, the likelihood function

of the observed data is

$$L(data) = \prod_{i=1}^{n} \left[ \int_{0}^{y_{i,2} \wedge y_{i,3}} f_1(x) f_2(y_{i,2} - x) f_3(y_{i,3} - x) dx \right]$$

$$= \prod_{i=1}^{n} \int_{0}^{y_{i,2} \wedge y_{i,3}} \left[ \frac{1}{\Gamma(\alpha_1)} \frac{1}{\beta^{\alpha_1}} x^{\alpha_1 - 1} \exp\{-\frac{x}{\beta}\} \right.$$

$$\times \frac{1}{\Gamma(\alpha_2)} \frac{1}{\beta^{\alpha_2}} (y_{i,2} - x)^{\alpha_2 - 1} \exp\{-\frac{y_{i,2} - x}{\beta}\}$$

$$\left. \times \frac{1}{\Gamma(\alpha_3)} \frac{1}{\beta^{\alpha_3}} (y_{i,3} - x)^{\alpha_3 - 1} \exp\{-\frac{y_{i,3} - x}{\beta}\} \right] dx$$

$$= \prod_{i=1}^{n} \left[ \frac{1}{\Gamma(\alpha_1)\Gamma(\alpha_2)\Gamma(\alpha_3)} \frac{1}{\beta^{\alpha_1 + \alpha_2 + \alpha_3}} \exp\{-\frac{1}{\beta}(y_{i,2} + y_{i,3})\} \right.$$

$$\left. \times \int_{0}^{y_{i,2} \wedge y_{i,3}} x^{\alpha_1 - 1} (y_{i,2} - x)^{\alpha_2 - 1} (y_{i,3} - x)^{\alpha_3 - 1} \exp\{\frac{x_i}{\beta}\} dx \right].$$

As before, the MLEs will have to be obtained numerically.

Let us consider the details of the EM-algorithm. The Gamma distribution is a member of the exponential family with sufficient statistics $X$ and $\log(X)$. For the two-layer tree, we need to compute just the conditional expectation of $X_1$ and $\log(X_1)$, the unknown delays on the first link. The conditional distribution $[X_1 | Y_1 = a, Y_2 = b]$ is now given by

$$(6) \qquad g(x) = \frac{x^{\alpha_1 - 1}(a - x)^{\alpha_2 - 1}(b - x)^{\alpha_3 - 1} \exp\{\frac{x}{\beta}\}}{C(a, b; \alpha_1, \alpha_2, \alpha_3, \beta)}, \quad 0 < x < a \wedge b,$$

where the proportionality constant is

$$C(a, b; \alpha_1, \alpha_2, \alpha_3, \beta) = \int_{0}^{a \wedge b} x^{\alpha_1 - 1}(a - x)^{\alpha_2 - 1}(b - x)^{\alpha_3 - 1} \exp\{\frac{x}{\beta}\} dx.$$

This can be used to compute $E[X_1|Y_1, Y_2]$ and $E[\log(X_1)|Y_1, Y_2]$ numerically. Note that

$$E[X_1|Y_1, Y_2] = \frac{C(Y_1, Y_2; \alpha_1 + 1, \alpha_2, \alpha_3, \beta)}{C(Y_1, Y_2; \alpha_1, \alpha_2, \alpha_3, \beta)}.$$

How well does this extend to trees with more than two layers? It turns out that the full MLE is still not feasible. However, a combination of "local" MLEs and a grafting idea (along the lines of [8]) is feasible. Consider the 3-layer tree in Figure 5. Suppose we use a flexicast experiment with 3 bicasts $\langle 4, 5 \rangle$, $\langle 6, 7 \rangle$, and $\langle 5, 6 \rangle$. The bicast scheme $\langle 4, 5 \rangle$ splits at node 2. So we can combine links 1 and 2 into a single link and use the previous results for the two-layer tree to get estimates for this subtree. Note that the delay distribution for the combined links 1 and 2 is $\Gamma(\alpha_1 + \alpha_2, \beta)$. So we can get "local" MLEs for $\alpha_1 + \alpha_2$, $\alpha_4$, $\alpha_5$ and $\beta$ from the bicast experiment $\langle 4, 5 \rangle$. Using a similar argument, we can get estimates for $\alpha_1 + \alpha_3$, $\alpha_6$, $\alpha_7$ and $\beta$ from the bicast scheme $\langle 6, 7 \rangle$ and estimates for $\alpha_1$, $\alpha_2 + \alpha_5$, $\alpha_3 + \alpha_7$ and $\beta$ from the bicast scheme $\langle 5, 6 \rangle$. Now we can use one of several methods to combine these estimates to get a unique set of estimates for all of the $\alpha_k$ and $\beta$. Possible methods include ordinary or weighted LS.

We do not pursue this strategy here as the specifics work only for special cases. The main message here is that it is not easy to compute the full MLE even in very simple cases, and the problem becomes completely intractable as the size of the tree and number of children in the links grow.

## 4.3. Method-of-moments estimation

We discuss the use of method of moments which estimates the parameters by matching the population moments to the sample moments using some appropriate loss function. General losses are possible, but squared error loss leads to more tractable optimization and the large-sample properties are easy to establish.

Let $H = \{1, \ldots, m\}$ be the index set of the probes used in the probing experiment. Denote the observed data $Y_r(1), \ldots, Y_r(m)$ as $Y_r(H)$. Let $M_i^j(H)$ be the observed $i$-th moment for the $j$-th scheme based on the probes in $H$. Let $\mathcal{M}_i^j(\theta)$ be the functional form of the $i$-th moment from the $j$-th probing scheme. For example, for the two-layer tree in Figure 1 with Gamma$(\alpha_k, \beta_k)$ distributions on each link, we get the following relationships:

$$\mathrm{E}(Y_2) = \alpha_1 \beta_1 + \alpha_2 \beta_2,$$
$$\mathrm{E}(Y_3) = \alpha_1 \beta_1 + \alpha_3 \beta_3,$$
$$\mathrm{Cov}(Y_2, Y_3) = \alpha_1 \beta_1^2,$$
$$\mathrm{E}(Y_2 - \nu_2)^2 (Y_3 - \nu_3) = 2\alpha_1 \beta_1^3,$$
$$\mathrm{Var}(Y_2) = \alpha_1 \beta_1^2 + \alpha_2 \beta_2^2,$$
$$\mathrm{Var}(Y_3) = \alpha_1 \beta_1^2 + \alpha_3 \beta_3^2.$$

We can now estimate the parameters by minimizing the squared error loss

$$Q(\theta; \mathbf{M}(H)) = \sum_{j=1}^{m} \sum_{i} \left[ M_i^j(H) - \mathcal{M}_i^j(\theta) \right]^2.$$

This is a special case of the nonlinear least squares problem and can be solved using iterative methods such as the Gauss-Newton procedure (see [1] for example). After rewriting the loss function as a single sum over all the moments, we consider the derivatives

(7)
$$\frac{\partial Q(\theta; \mathbf{M}(H))}{\partial \theta_j} = -2 \sum_{i} [M_i(H) - \mathcal{M}_i(\theta)] \frac{\partial \mathcal{M}_i(\theta)}{\partial \theta_j}.$$

These can be expressed in matrix form as

$$\left[ \frac{\partial Q(\theta; \mathbf{M}(H))}{\partial \theta} \right] = D'[M(H) - \mathcal{M}(\theta)],$$

where $D_{i,j} = \frac{\partial \mathcal{M}_i(\theta)}{\partial \theta_j}$. The moments at the true value can be expanded using a Taylor series expansion around an initial guess $\theta^{(0)}$ as $\mathcal{M}(\theta_0) \approx \mathcal{M}(\theta^{(0)}) + D(\theta_0 - \theta^{(0)})$. Computing the residuals and replacing the true value with the observed moments gives an updating scheme based on solving a linear system. Thus at iteration $q$, we have the following linear system.

$$M(H) - \mathcal{M}(\theta^{(q)}) = D\beta.$$

Solving this, we get the next iteration as $\theta^{(q+1)} = \theta^{(q)} + \hat{\beta}$.

In general, each iteration should be closer to the minimizer. However, there can be situations where the step increases the sum of squares. To avoid this, we recommend the modified Gauss-Newton in which the next iteration is given by $\theta^{(q+1)} = \theta^{(q)} + r\hat{\beta}$ where $0 < r \leq 1$. This fraction can be chosen adaptively at each step. If the full step reduces the sum of squares, then it is taken. Otherwise, we can set $r = .5$. If the half step fails to reduce the sum of squares, then it can be halved again. This guarantees that the loss function is reduced with every step and gives convergence to a stationarity point. Examination of the derivatives will indicate if the point is a minimum.

The algorithm has useful complexity properties in terms of both memory and computation. Since the estimation is based only on the moments, these values are all that need to be stored. This is a vast improvement over algorithms that require all of the data or the counts of the binned data. Further, the efficient implementation of the algorithm, involving a QR factorization and one matrix inversion, gives computational complexity of $\mathcal{O}(m^3)$ where $m$ is the number of required moments. Again, this is a large improvement over other methods that have exponential complexity. Further improvement is achieved in many cases using sparse matrix techniques. These two points make the approach ideal for application requiring real-time estimates.

The ordinary non-linear least-squares (OLS) method of moment (MOM) estimators can be inefficient as the different moments are correlated and have unequal variances. Since these can often be computed and estimated easily, one can use generalized least-squares (GLS) to improve the efficiency. A limited comparison of the efficiencies is given in the next subsection.

It is easy to show that the method-of-moment estimators based on OLS or GLS are consistent and asymptotically normal as the sample sizes (numbers of probes) increase on a given network. The large-sample distribution can be used to compute approximate confidence regions and hypothesis tests which are useful in monitoring applications.

### *4.4. Relative efficiency of the method-of-moments: a limited study*

We conducted a small simulation study to assess the performance of the MOM estimators versus the MLE. This was done on a two-layer binary tree (left panel of Figure 1) for exponential distributions. This is one of the few instances when it is practical to construct the MLE. We used two MOM estimators: the OLS and the GLS schemes (described above). The GLS methods used a weighting scheme based on an empirical estimate of the covariance of the observed moments. Relative efficiency is defined as the ratio of the variance of the MLE to the variance of the estimator of interest.

We considered two cases: a) each link has the same mean; i.e., $1/\lambda_k = 1/2$, $k = 1, 2, 3$, and b) each link has its own mean, $1/\lambda_1 = 1/2$, $1/\lambda_2 = 1/4$, and $1/\lambda_3 = 1/6$, respectively. For both scenarios, 1000 data sets of size 3000 were generated. Boxplots of the three sets of estimators are shown in Figures 6 and 7. The procedures appear to be unbiased. When the means are the same, the relative efficiencies of the OLS MOM are 1.72, 1.33, 1.41 and the relative efficiencies of the GLS MOM are 1.12, 1.11, and 1.12. When the means are different, the relative efficiencies for the OLS are 2.43, 5.09, and 9.13 and the relative efficiencies for the GLS are 1.07, 1.24, and 1.34.

In this example, the GLS MOM appears to be quite efficient compared to the MLE. However, a much more extensive study is clearly needed to quantify the performance of the MOM estimators.

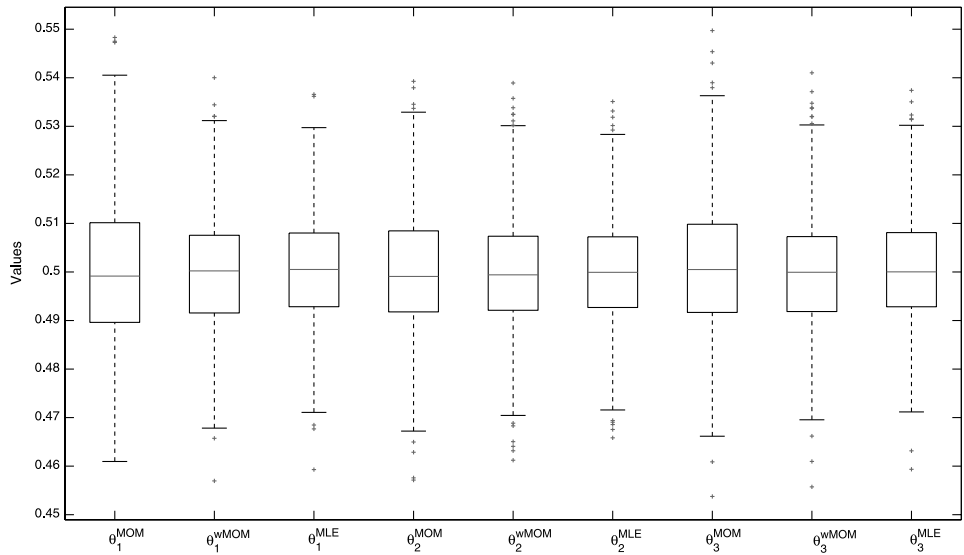

FIG 6. *Boxplots comparing the MLE with MOM: Case 1 – All means are equal.*

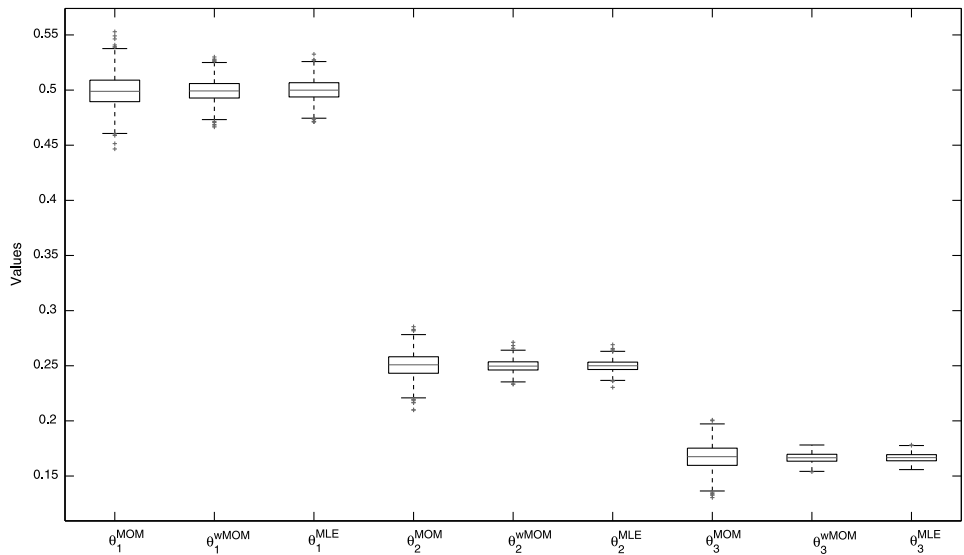

FIG 7. *Boxplots comparing the MLE with MOM: Case 2 – The three means are different.*

## *4.5. Analysis using the NS-2 simulator*

We now describe a study of the MOM estimators in a simulated network environment. The *ns-2* package is a discrete event simulator that allows one to generate network traffic and transmit packets using various network transmission protocols, such as TCP, UDP, ([15]) over wired or wireless links. The simulator allows the underlying link delays to exhibit both spatial and temporal dependence with correlation between sister links (children of the same parent, *i.e.* 4, 5, and 6.) around .25 and autocorrelation about .2 on all of the links. Thus, we can study the performance of the active tomography methods under more realistic scenarios.

We used the topology shown in Figure 8 with a multicast transmission scheme. The capacity of all links was set to the same size (100 Mbits/sec), with 11 sources (10 TCP and one UDP) generating background traffic. The UDP source sent 210 byte long packets at a rate of 64 kilobits per second with burst times exponentially distributed with mean .5s, while the TCP sources sent 1,000 byte long packets every .02s. The main difference between these two transmission protocols is that UDP transmits packets at a constant rate while TCP sources linearly increase their transmission rate to the set maximum and halve it every time a loss is recorded. The length of the simulation was 300 seconds, with probe packets 40 bytes long injected into the network every 10 milliseconds for a total of about 3,000. Finally, the buffer size of the queue at each node (before packets are dropped and losses recorded) was set to 50 packets.

We studied only the continuous component of the delay distribution, i.e., the portion of the path-level data that contain zero or infinite delays was removed. The traffic-generating scenario described above resulted in approximately uniform waiting times in the queue (see Figure 9). This is somewhat unrealistic in real network situations where traffic tends to be fairly bursty [11], but it provides a simple scenario for our purposes. Estimating the unknown parameters for this model is equivalent to estimating the maximum waiting time for a random packet.

Figure 10 shows quantile-quantile plots using simulated values from the fitted distributions versus the observed *ns-2* delays for both the links and paths. Specifically, we estimated the parameters for the uniform distributions using the moment estimation procedure and then generated data based on these parameter values. The fitted values were: $\hat{b} = [.89, .79, .53, 1.10, 1.09, 1.13]$. The estimation procedure does quite well on all of the links except for the interior link 3. The algorithm seems to compensate for this under-estimation (about 40%) by slightly overestimating the parameters for each of the descendants of link 3, as evidenced by the closely matched quantiles for the end-to-end data. This error is probably the result of several factors. First, link 3 deviates the most from the uniform distribution with the last bin in Figure 9 being too thin. Secondly, the algorithm appears to be moderately affected by the violations of the independence assumptions, particularly the spatial dependence among the children of link 3. This could likely be somewhat relieved by using a larger sample size and accounting for the empty queue probabilities. Nevertheless, the estimation performs well overall.

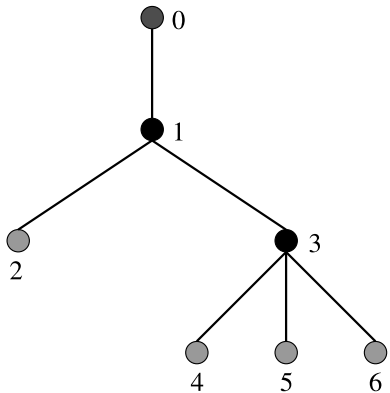

FIG 8. *Portion of the UNC network used in the ns-2 simulaton scenario.*

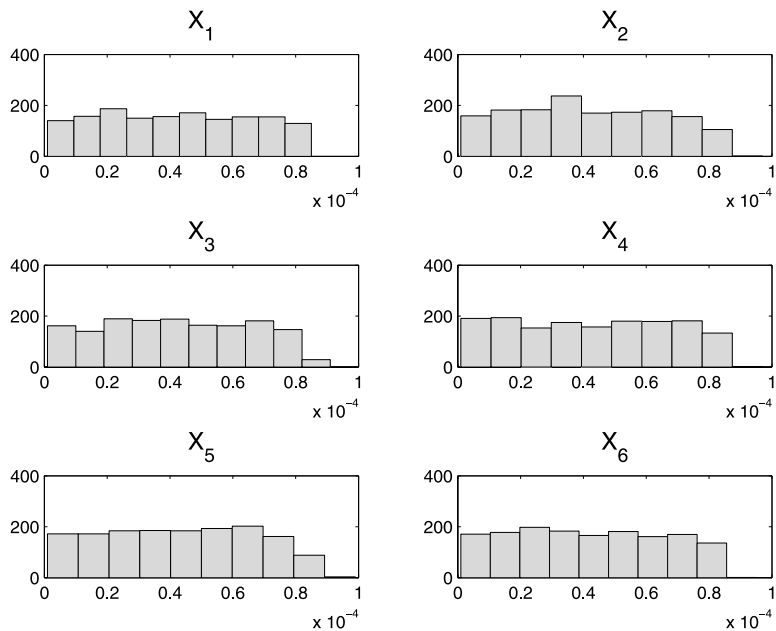

Fig 9. *Histogram of link delays for the* ns-2 *simulation.*

## 5. Summary

There are a number of other interesting problems that arise in active network tomography. There are usually multiple source nodes, which raises the issue of how to optimally design flexicast experiments for the various sources. We have also assumed that the logical topology of the tree is known. However, only partial knowledge of the network topology is typically available, and one would be interested in using the path level information to *simultaneously* discover the topology and estimate the parameters of interest. The topology discovery problem is computationally difficult (NP-hard), but methods using a Bayesian formulation as well as those based on clustering ideas have been proposed in the literature (see [3] and references therein).

Active tomography techniques are useful for monitoring network quality of service over time. However, this application requires that path measurements are collected sequentially over time and appropriately combined. The probing intensity, the type of control charts to be used for monitoring purposes, and the use of path- vs link- information are topics of current research.

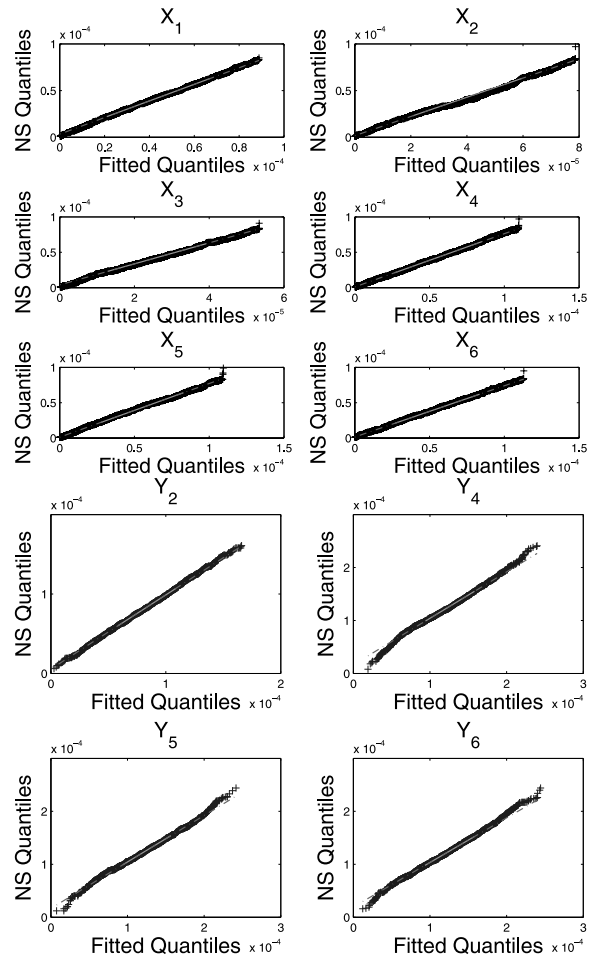

FIG 10. *QQ plots for the links and paths of the* ns-2 *simulator example. The fitted quantiles come from simulating delays from distributions with parameters given by the fitted values of the data.*

**Acknowledgments.** The authors are grateful to: Jim Landwehr, Lorraine Denby, and Jean Meloche of Avaya Labs for making their ExpertNet tool available for VoIP data collection and for many useful discussions on network monitoring; Jim Gogan and his team from the IT Division at UNC for their technical support in deploying the ExpertNet tool on their campus network, for troubleshooting hardware problems and for providing information about the structure and topology of the network; Don Smith of the CS Department at UNC for helping us establish the collaboration with the UNC IT group; and to two referees for several useful comments that have improved the paper.

## References

[1] BATES, D. AND WATTS, D. (1988). *Nonlinear Regression Analysis and Its Applications.* Wiley, New York. MR1060528
[2] CÁCERES, R., DUFFIELD, N. G., HOROWITZ, J. AND TOWSLEY, D. F. (1999). Multicast-based inference of network-internal loss characteristics. *IEEE Trans-*

actions on Information Theory **45** 2462–2480. MR1725131
[3] CASTRO, R., COATES, M., LIANG, G., NOWAK, R. AND YU, B. (2004). Network tomography: Recent developments. *Statist. Sci.* **19** 499–517. MR2185628
[4] CHEN, A., CAO, J. AND BU, T. (2005). Network tomography: Identifiability and fourier domain estimation. Submitted.
[5] COATES, M. J. AND NOWAK, R. D. (2000). Network loss inference using unicast end-to-end measurement. In *Proceedings of the ITC Conference on IP Traffic, Modelling and Management.*
[6] LAWRENCE, E. (2005). Flexicast network delay tomography. Ph.D. thesis, University of Michigan.
[7] LAWRENCE, E., MICHAILIDIS, G. AND NAIR, V. N. (2003). Maximum likelihood estimation of internal network link delay distributions using multicast measurements. In *Proceedings of the 37th Conference on Information Sciences and Systems.*
[8] LAWRENCE, E., MICHAILIDIS, G. AND NAIR, V. N. (2006). Network delay tomography using flexicast experiments. *J. Roy. Statist. Soc. Ser. B* **68** 785–813. MR2301295
[9] LIANG, G. AND YU, B. (2003). Maximum pseudo likelihood estimation in network tomography. *IEEE Transactions on Signal Processing* **51** 2043–2053.
[10] LO PRESTI, F., DUFFIELD, N. G., HOROWITZ, J. AND TOWSLEY, D. (2002). Multicast-based inference of network-internal delay distributions. *IEEE Transactions on Networking* **10** 761–775.
[11] PARK, K. AND WILLINGER, W. (2000). *Self-Similar Network Traffic and Performance Evaluation.* Wiley Interscience, New York.
[12] SHIH, M.-F. AND HERO, A. O. (2003). Unicast-based inference of network link delay distributions with finite mixture models. *IEEE Transactions on Signal Processing* **51** 2219–2228.
[13] TSANG, Y., COATES, M. AND NOWAK, R. D. (2003). Network delay tomography. *IEEE Transactions on Signal Processing* **51** 2125–2135.
[14] VARDI, Y. (1996). Network tomography: Estimating source-destination traffic intensities from link data. *J. Amer. Statist. Assoc.* **91** 365–377. MR1394093
[15] WALRAND, J. (2002). *Communications Networks: A First Course.* McGraw-Hill.
[16] XI, B., MICHAILIDIS, G. AND NAIR, V. N. (2003). Least squares estimates of network link loss probabilities using end-to-end multicast measurements. In *Proceedings of the 37th Conference on Information Sciences and Systems.*
[17] XI, B., MICHAILIDIS, G. AND NAIR, V. N. (2006). Estimating network loss rates using active tomography. *J. Amer. Statist. Assoc.* **101** 1430–1438. MR2279470
[18] ZHANG, C.-H. (2005). Estimation of sums of random variables: Examples and information bounds. *Ann. Statist.* **33** 2022–2041. MR2211078